\documentclass[aps,pre,twocolumn,amssymb,showpacs,groupedaddress]{revtex4-1}
\usepackage{graphicx}
\usepackage{amsmath}

\usepackage{changes}
\def\beq{\begin{equation}}
\def\eeq{\end{equation}}
\def\bea{\begin{eqnarray}}
\def\eea{\end{eqnarray}}

\begin{document}



\title{Measuring geometric frustration in twisted inextensible filament bundles 
}


\author{Andreea Panaitescu,$^1$ Gregory M. Grason,$^2$ and Arshad Kudrolli$^{1,}$}
\email[]{AKudrolli@clarku.edu}
\affiliation{$^1$Department of Physics, Clark University, Worcester, MA 01610,\\ 
$^2$Department of Polymer Science and Engineering, University of Massachusetts, Amherst, MA 01003}

\date{\today}

\begin{abstract}
We investigate with experiments and novel mapping the structure of a hexagonally ordered filament bundle that is held near its ends and progressively twisted around its central axis. The filaments are free to slide relative to each other and are further held under tension-free boundary conditions. Measuring the bundle packing with micro x-ray imaging, we find that the filaments develop the helical rotation $\Omega$ imposed at the boundaries. We then show that the observed structure is consistent with a mapping of the filament positions to disks packed on a dual non-Euclidean surface with a Gaussian curvature which increases with twist. We further demonstrate that the mean inter-filament distance is minimal on the surface which can be approximated by a hemisphere with an effective curvature $K_{eff} = 3\Omega^2$. Examining the packing on the dual surface, we analyze the geometric frustration of packing in twisted bundles and find the core to remain relatively hexagonally ordered with inter-filament strains growing from the bundle center, driving the formation of defects at the exterior of highly twisted bundles. 
\end{abstract}


\maketitle


\section{Introduction}
Twisted bundles of filaments are widely encountered in ropes, yarns, animal tissue, and bacterial flagella~\cite{costello90,pan02,bruss12,haines16,liem00,macnab77}. 
 A twisted geometry has been shown to lead to strengthening in materials ranging from textiles to carbon nanotube bundles~\cite{hearle69,zhang04}, rich timber in spider silk~\cite{osaki12,hennecke13}, and resonances in photonic crystal fibers~\cite{wong12}. Despite their broad use, the organization 
of constituent quasi-1D filaments in twisted bundles remain a long-standing and unsolved problem, even for the simplest case of filaments with uniform circular cross sections.  Long appreciated by textile scientists~\cite{schwarz52,hearle69,pan02}, the packing problem can be viewed from the perspective of a 2D planar transect through the bundle, i.e., parallel, untwisted bundles appear as constant diameter circle packings, and hence, permit uniform, hexagonal close packing. In contrast, application of twist inclines filaments with respect to the plane, leading to planar sections filled with apparently non-circular elements~\cite{neukrich02,olsen10,bohr11}, whose shape and orientation vary throughout the packing~\cite{grason15}.  

From the geometric perspective of packing variable-shape elements in the 2D plane, it is intuitive that imposing twist to an initially close-packed array requires the packing to deform to avoid overlaps.  Far less obvious is how precisely the local and global features of the packing evolve with progressively increasing twist. 
Spacing between quasi-1D filaments is characterized by distance of closest separation, dependent on both shape and orientation of filaments.   The local constraints of non-overlap are easily formulated in the plane normal to a given filament: center-to-center separation to all neighbors must not be less than the diameter, $d$, in this plane. Without resorting to variable shape elements, the relative variation of the reference plane and the local filament orientation throughout the bundle makes it impossible to represent the contact structure of the entire bundle simply in that plane. 

Remarkably, a recently developed approach shows that inter-filament distances are instead more straightforwardly represented by mapping filament positions onto a non-Euclidean surface~\cite{bruss12, grason15}. This dual surface has azimuthal symmetry and a positive Gaussian curvature proportional to the square of bundle twist.  The geodesic distances between points on the dual surface are equivalent to separations between corresponding helical curves in the bundle. Thus, the packing of constant-diameter disks on this dual surface properly encodes both the local and global constraints imposed by non-overlap in 3D twisted, multi-filament bundles within a single 2D manifold.  To date, this geometric mapping has been exploited to understand complex patterns of topological defects~\cite{grason12,amir12,bruss13} and morphological selection~\cite{hall16} favored in models of cohesive ground-states of twisted bundles. However, a physical demonstration of a filament bundle which actually tests the underlying assumptions of the geometric mapping of inter-filament contact to our knowledge has never previously been accomplished. 

Here, we develop new experiments to understand the evolution of a nominally hardcore repulsive  bundle with twist imposed and test the mapping equivalence by performing x-ray tomography of the filament shapes and positions throughout the 3D structure.  Rather than groundstate or equilibrium structure, our interest is in the collective deformation of the filament packing as the bundle is progressively twisted starting from a parallel, closed-packed hexagonal arrangement. We first show that the specific pattern of collective deformation in twisted bundles is a consequence of the non-trivial constraints on the inter-filament spacing imposed by non-parallel orientation of the filaments. We then demonstrate that the mapping of filament positions onto the quasi-hemispherical surfaces accurately encodes the inter-filament contact distances.  These experiments verify that mechanically-imposed longitude twist of a bundle introduces {\it geometric frustration} to the lateral packing of filament, require non-equal spacing of filaments for any finite measure of twist.  

\section{Experimental System}
\begin{figure}
\begin{center}
\includegraphics[width=0.35\textwidth]{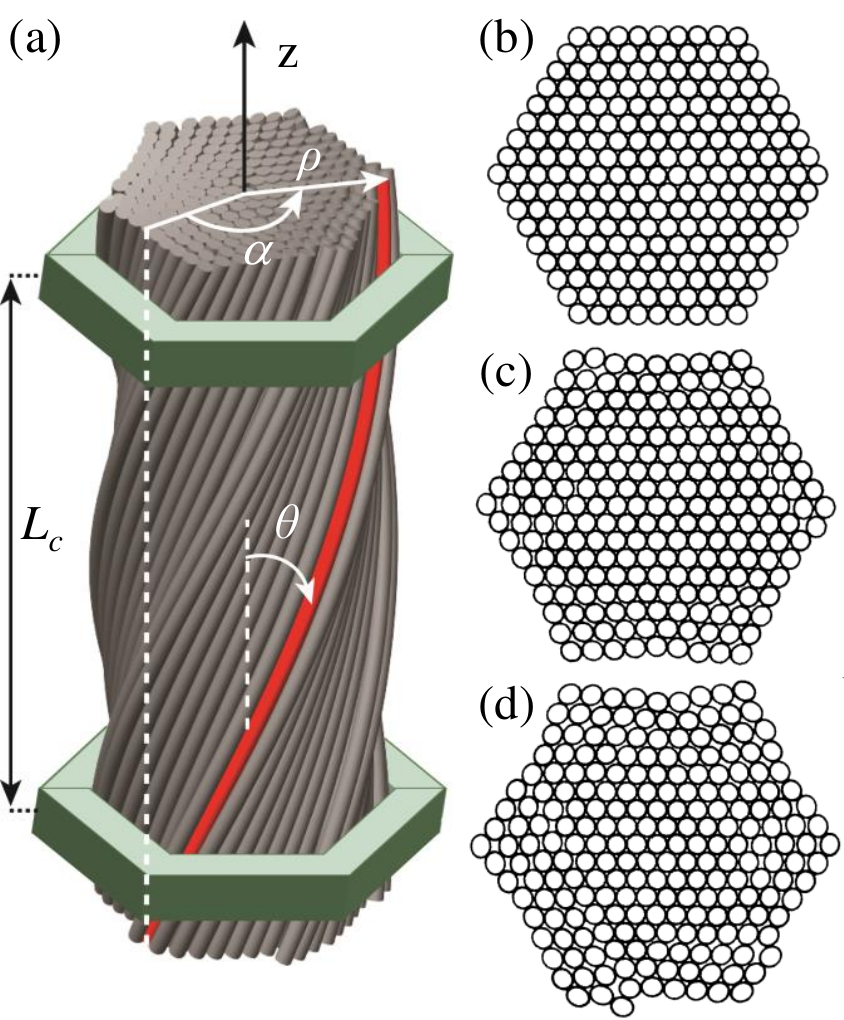}
\end{center}
\caption{(a) A schematic of a fiber bundle held within two hexagonal clamps that are twisted through a prescribed angle $\alpha$. Bundle transects obtained with x-ray scanning corresponding to (b) $\alpha = 0^\circ$, (c) $\alpha = 129^\circ$, and (d) $\alpha = 176^\circ$.}
\label{fig:experimental}
\end{figure}

A schematic of the experimental system is shown in Fig.~\ref{fig:experimental}(a). The bundle consists of hollow polypropylene rods arranged initially in a hexagonal lattice such that the bundle sides corresponds to nine filaments. The filaments have circular cross sections with diameter $d = 3.1 \pm 0.06$\,mm, thickness $t = 0.2 \pm 0.01$\,mm, length $L = 150$\,mm. The bending or flexural modulus of the filament is measured with a standard 3-point bending test apparatus and found to be $3.3$\,GPa. The stretching modulus of the filament used is immeasurably high on this scale and therefore the filaments can be treated as inextensible.  The bundle is held together near its ends by two clamps with a hexagonal cross section in which the filaments fit snuggly. The clamps are cast out of vinylpolysiloxane with Young modulus of $384$\,kPa which allows the bundle cross section to expand as it is twisted. The clamps, separated by distance $L_c/d = 36$, are then further mounted inside rigid circular end caps of a circular cylinder. A twist is applied by keeping the bottom fixed and rotating the top clamp through a prescribed angle $\alpha$ as shown in Fig.~\ref{fig:experimental}(a). The filaments are allowed to slide freely relative to each other as they bend by adding a layer of talc. Thus, the filaments do not experience any significant stretching along their length as no tension is applied at the ends. 
 
\begin{figure}
\begin{center}
\includegraphics[width=0.4\textwidth]{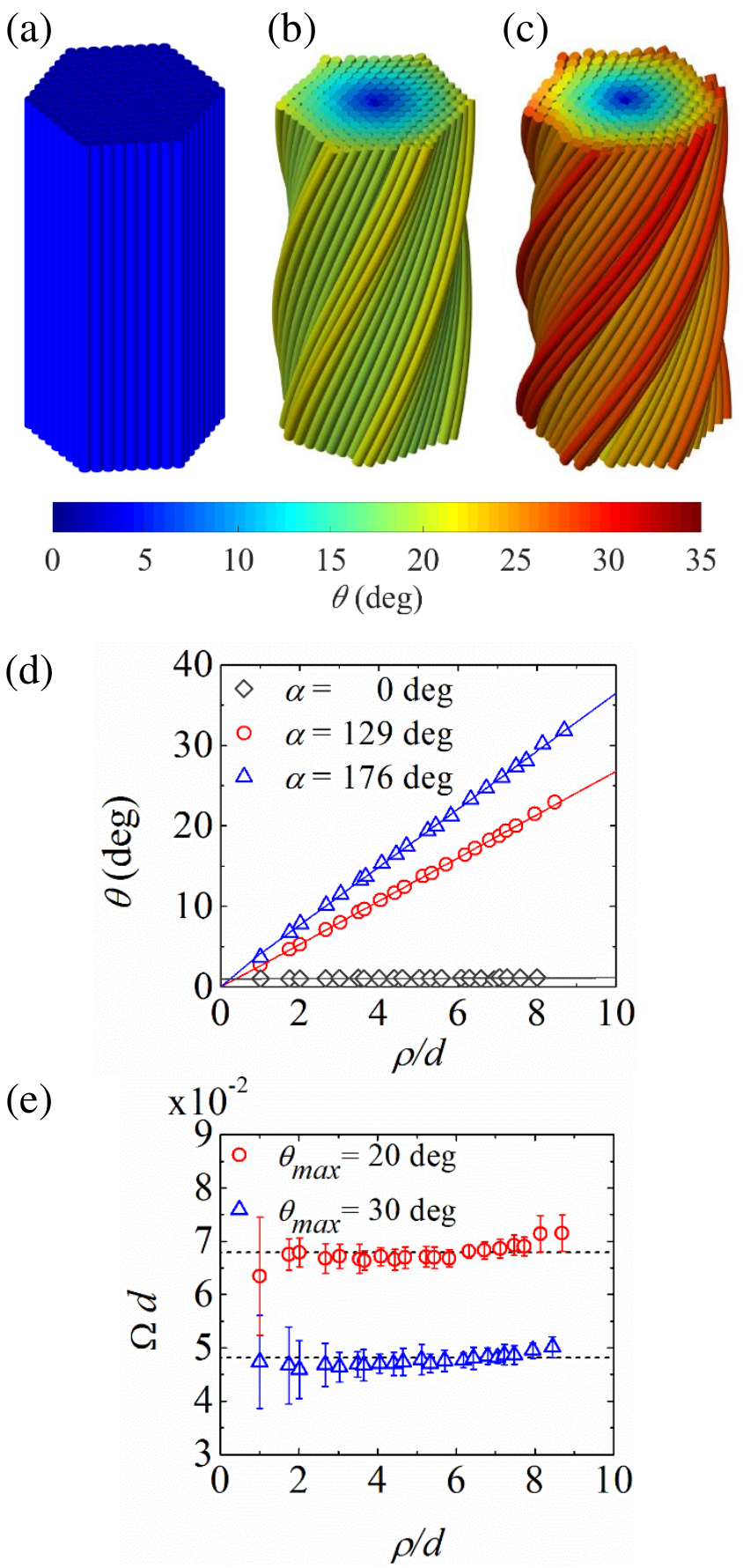}
\end{center}
\caption{(a-c) Reconstructed filament bundles color coded according to the tilt angle $\theta$ of the filament shown in the color bar. The maximum tilt angle  $\theta_{max} = 0^\circ$ (a), $\theta_{max} = 20^\circ$ (b), and $\theta_{max} = 30^\circ$ (c).  (d) $\theta$ as a function of the helical radius $\rho$ binned in $d/2$ intervals. The lines are linear fits. 
(e) The average helical rotation $\Omega$ as a function of the helical radius $\rho$ binned in $d/2$ intervals. The bars correspond to the root mean square of $\Omega$ measured for filaments in that bin, and the horizontal dashed lines to the mean values $\Omega_m$. 
}
\label{fig:experimental2}
\end{figure} 

\section{Measurement of bundle structure}

\begin{figure*}
\begin{center}
\includegraphics[width=0.7\textwidth]{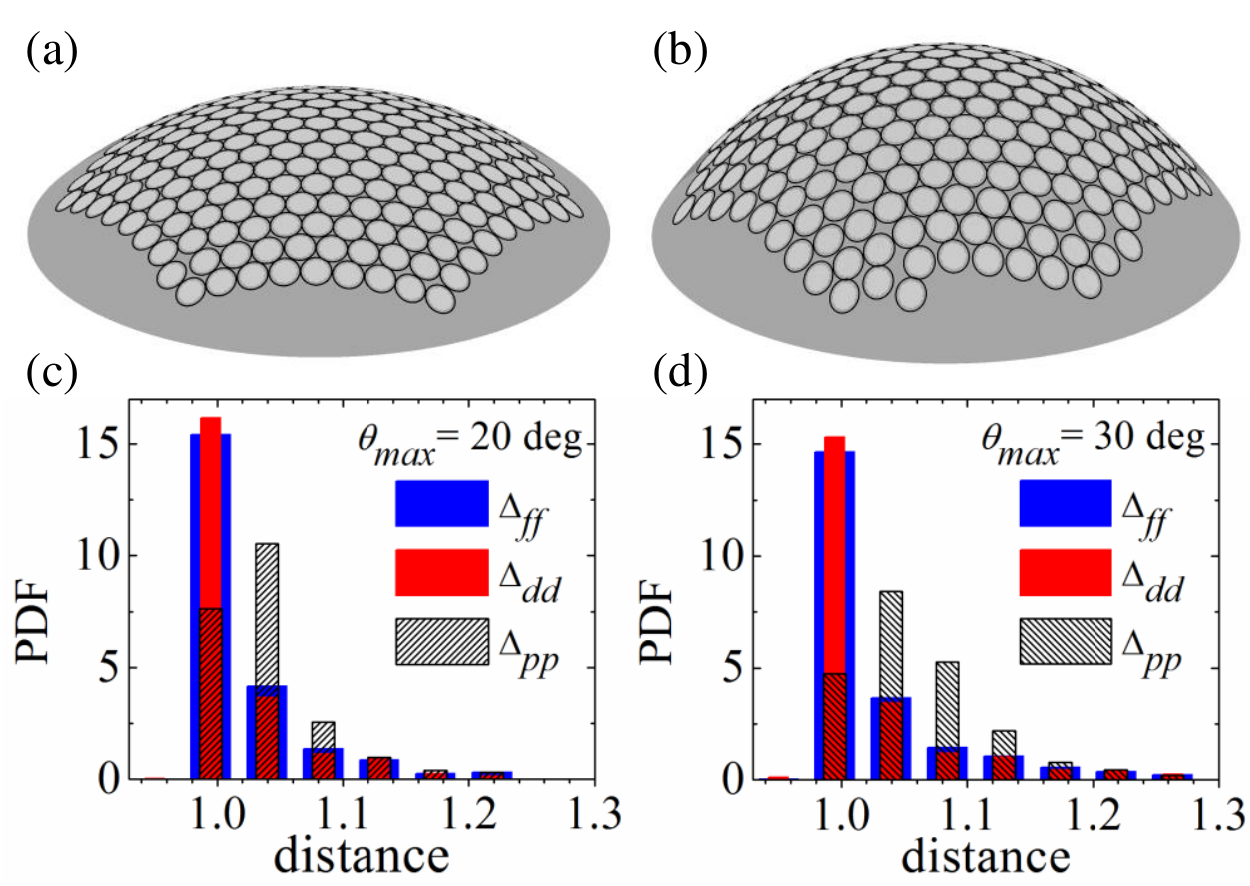}
\end{center}
\caption{Disks plotted on a surface given by Eq.~\ref{mapping}  for (a) $\theta_{max} = 20^{\circ}$ and (b) $\theta_{max} = 30^{\circ}$. (c,d) The probability distribution function (PDF) of  distances shown in the legend. $\Delta_{ff}$ and  $\Delta_{dd}$ are observed to be similar, and distinct from the PDF of $\Delta_{pp}$ for both (c) $\theta_{max} = 20^{\circ}$ and (d) $\theta_{max} = 30^{\circ}$.}
\label{fig:distribution}
\end{figure*}

We probe the internal structure of the fiber bundle with a Varian Medical Systems micro focus x-ray Computed Tomography instrument by scanning a central region of the bundle away from the clamps. Each scan consists of 148 transects equally spaced along the bundle axis $z$ and separated by $0.0226d$. Fig.~\ref{fig:experimental}(b) shows a sample transect before twist is applied, and Fig.~\ref{fig:experimental}(c-d) show examples  for progressively higher $\alpha$. While the filament cross sections appear circular before twist is applied, it can be noted that they appear increasing elliptical as the filaments tilt in response to twist. To quantify the position and orientation of individual fibers, we first locate the intersection of each filament in a horizontal slice using a segmentation algorithm implemented in MATLAB. Then, the cross section of each filament is fitted to an ellipse to extract its center, and then tracked from one slice to the next to determine the coordinates of its central axis~\cite{yadav13}. These points 
are then fitted with a helix characterized by its radius $\rho$ and helical rotation $\Omega$. 

We then use those parameters to reconstruct the fiber bundle corresponding to the length between the clamps in Figs.~\ref{fig:experimental2}(a-c). Each filament is denoted with a color corresponding to the angle $\theta$ subtended by the filament with the bundle axis according to the color bar. Fig.~\ref{fig:experimental2}(d) shows a plot of $\theta$ measured 
 as a function of $\rho$ binned in $d/2$ intervals. For a given imposed $\alpha$, we observe that $\theta$ increases linearly from 0 at the center of the bundle to the filaments furthest from the center. In the case of the three bundles shown in Figs.~\ref{fig:experimental2}(a-c), we measure the angle of tilt of the 48 filaments in the outer most layer and find the maximum tilt angle  $\theta_{max} = 0^\circ$, $20.5 \pm 1.4^\circ$, and $28.8 \pm 1.9^\circ$, respectively. For simplicity of annotation, we round up the values of these angles to be $\theta_{max} = 0^\circ$, $20^\circ$, and $30^\circ$. 

We plot the measured helical rotation $\Omega$  as a function of $\rho$ binned in $d/2$ intervals in Fig.~\ref{fig:experimental2}(e). We find that $\Omega$ is indeed constant for each imposed twist with mean helical rotation $\Omega_{m}d = (5.0 \pm 0.6) \times 10^{-2}$ for $\theta_{max} = 20^{\circ}$, and $\Omega_m d = (6.8 \pm 0.6) \times 10^{-2}$ for $\theta_{max} = 30^{\circ}$. To understand these mean values, one can calculate the helical rotation by assuming that all filaments in the bundle are twisting through the same angle $\alpha$ over the length $L_c$. Then, $\Omega = \alpha/L_c$. This implies that $\Omega d = 4.7 \times 10^{-2}$ and $7.4 \times 10^{-2}$ for $\alpha = 129^\circ$ and $\alpha = 176^\circ$, respectively. These values can be noted to be in agreement with the measured values within experimental error. 

Thus, one concludes that the filaments self-organize to have roughly the same helical rotation, implying that filaments maintain neighbor contacts along their length, which is presumably favored in a compact bundle. An important assumption the mapping model~\cite{bruss12} is that $\Omega$ for the twisted filament bundle is constant and the structure in any bundle transect is the same as in any other transect to within a rotation. Our system gives at least one example of an experimental realization where this assumption is true.    

\section{Dual non-Euclidean Surface}

To understand the disruption of hexagonal packing in the 2D planar section with twist, we now analyze the bundle using the mapping to the non-Euclidean surface given by~\cite{grason15}:
\begin{equation}
\label{mapping}
\textbf{X}(\rho, \phi) = \rho \cos \theta (\rho)\big(\cos\phi \hat{x} + \sin\phi \hat{y}\big) + z(\rho) \hat{z}\,,
\end{equation}
where $\hat{x}$, $\hat{y}$ and $\hat{z}$ are Cartesian directions, $\phi$ is the azimuthal coordinate, and 
\begin{equation}
\label{eq: the}
\cos \theta (\rho) = 1/\sqrt{1+(\Omega \rho)^2}.
\end{equation}
The radial distance $\rho$ is mapped to an arc-distance measured from the top of ``dome-like" surface, with a profile that satisfies 
\begin{equation}
\label{eq: dzdrho}
\partial z/\partial \rho = - \sqrt{1-\cos^6 \theta(\rho)},
\end{equation}
which can be integrated to further obtain $z(\rho)$ in Eq.~\ref{mapping}. From Eqs. (\ref{eq: the}) and (\ref{eq: dzdrho}) it is straightforward to show for small-$\rho$ (near to the center of the bundle) $z(\rho) \simeq -\sqrt{3}/2 | \Omega| \rho^2$, such that the radius of curvature at $\rho \to 0$ is $r_s(\rho \to 0) = |d^2 z/d \rho^2|^{-1} =  | \Omega|/\sqrt{3}$, or equivalently the dual surface has a positive Gaussian curvature $K_{eff} = 3  \Omega^2$ at its center.  Intuitively, the positive curvature of this surface can be linked to the fact that tilted outer filaments subtend a larger distance along the azimuthal distance (by a factor $1/\cos \theta (\rho)$) than they would if not tilted.  Like the shortening of ``latitudes" a distance $\rho$ from the pole of a globe, there is increasing ``less room" available from filaments placed at further $\rho$ from the bundle center than there would be in a parallel bundle.

 Fig.~\ref{fig:distribution}(a) and Fig.~\ref{fig:distribution}(b) show the surfaces corresponding to $\theta_{max} = 20^0$ and $\theta_{max} = 30^0$. The tangent planes of this curved surface match the distance constraints of packing helical filaments in the bundle normal to their backbone orientation. Hence, equal diameter $d$ disks placed at mapped filament positions represent the contact structure of circular filaments of the same diameter, consistent with the apparent close-contact of neighboring disks on both surfaces.  We first quantitatively test the distance representation of the non-Euclidean surface, and then exploit the mapping to analyze and understand features of the deformation pattern.

\section{Testing the packing equivalence}

The equivalence between the twisted bundle packing and the disk packing on a curved surface presumes that the minimum distance between two nearest neighbor filaments in the bundle determined by center-to-center packing $\Delta_{ff}$ maps onto the geodesic distance normalized by the filament diameter, $\Delta_{dd}$, measured between equivalent points on a 2D surface of specific curved geometry.
We first test this quantitatively by comparing the probability distribution function (PDF) of $\Delta_{ff}$ and PDF of $\Delta_{dd}$ (where both distances are normalized by the filament diameter) in Fig.~\ref{fig:distribution}(c) for $\theta_{max} = 20^0$ and in Fig.~\ref{fig:distribution}(d) for $\theta_{max} = 30^0$. In calculating $\Delta_{dd}$, we assume the specific surface shape predicted by Eq. (\ref{mapping}). In both cases, we find that the PDFs match within experimental errors, demonstrating that the perpendicular distance between filaments indeed correspond to a mapped packing of disks on the non-Euclidean surface. To contrast with the distributions obtained if the distance between the filaments is measured in the planar transect $\Delta_{pp}$, we observe that those PDFs are distinct. One clearly observes that PDF of $\Delta_{pp}$ is qualitatively shifted to larger separations, indicating that this perspective fails to provide an accurate measure of true inter-filament spacing.

\begin{figure}
\begin{center}
\includegraphics[width=0.45\textwidth]{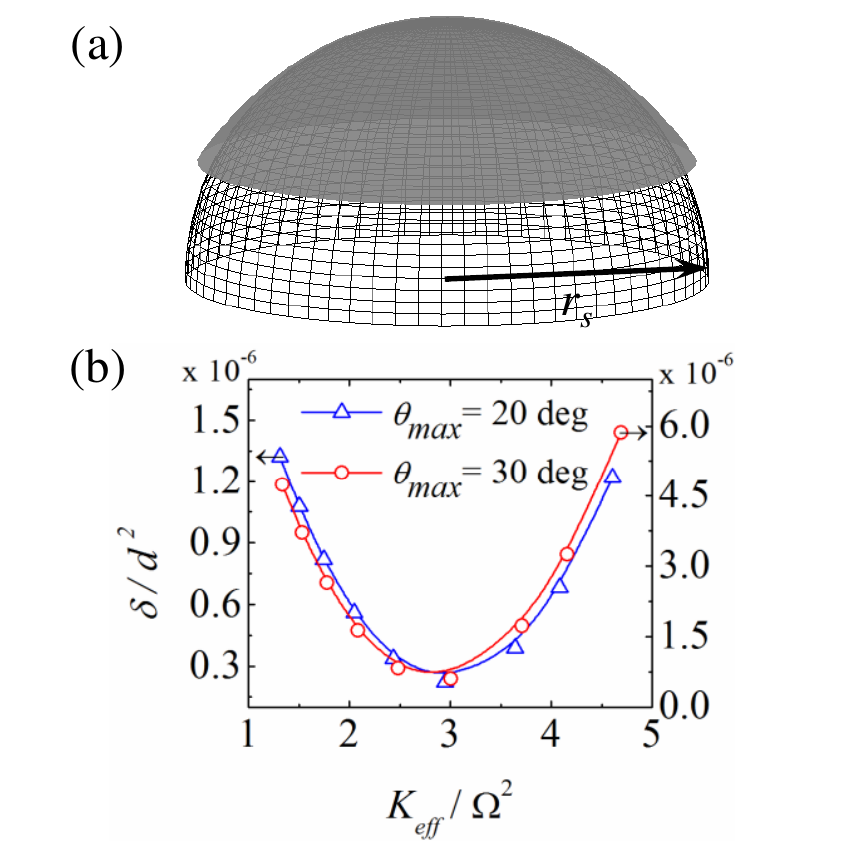}
\end{center}
\caption{(a) Comparison of Eq.~\ref{mapping} (solid surface) with spherical approximation for $\theta_{max} = 30^{\circ}$ (mesh) for $\theta_{max} = 30^{\circ}$. (b) $\delta^2$ as a function of the Gaussian curvature $K_{eff}$ is observed to reach a minimum when $K_{eff}/\Omega^2 = 3$ for both twist angles consistent with the quasi-hemispherical mapping.}
\label{fig:distribution2}
\end{figure}

Next, we confirm that the shape given by the surface in Eq.~(\ref{mapping}) provides the optimal accuracy in describing inter-filament spacing.  Specifically, we consider mapping filament positions to surfaces of variable positive Gaussian curvature.  We further confirm that the surface given by Eq.~\ref{mapping} indeed corresponds to the minimum separation between the filaments by mapping them on to surfaces which have smaller as well as greater Gaussian curvature. To simplify the calculations, we approximate the non-Euclidean surface given by Eq.~\ref{mapping} with sphere of radius $r_{s}=K_{eff}^{-1/2}$ as shown in Fig.~\ref{fig:distribution2}(a). We map filament centers to spheres of variable radii $r_s$ and calculate the agreement of inter-element spacing through the parameter 
\begin{equation}
\delta^2(r_s) = \frac{1}{N} \sum_{i=1}^{N} N_i^{-1}\sum_{j=1}^{N_i} (\Delta_{ff}^{ij} - \Delta_{dd}^{ij})^2\,,
\end{equation}
where $j$ labels each of the $N_i$ neighbors of $i$th filament in the bundle, $N$ is the total number of filaments in the bundle and $\Delta_{\rm dd}^{ij}$ is determined according to the geodesic distances on the sphere of radius $r_s$. We plot  $\delta^2$ as a function of the Gaussian curvature $K_{eff} = r_s^{-2}$ of the corresponding sphere in Fig.~\ref{fig:distribution2}(b) for each $\theta_{max}$.  Here, $K_{eff}$ is scaled by the corresponding $\Omega$ in order to collapse the data on to a single curve. We observe that $\delta^2$ has a minimum at $K_{eff}/\Omega^2 = 3$, in precise agreement with the Gaussian curvature predicted by the geometric mapping calculated and reported in Ref.~\cite{grason15}. This demonstrates that not only does the mapping of filaments to a positively-curved surface provide a quantitatively more accurate description of inter-filament spacing than the naive view in the planar cut, but that the specific relationship between bundle twist $\Omega$ and positive surface curvature implied by Eq.~(\ref{mapping}) is required to capture the metric constraints in helically twisted, multi-filament packings.

\section{Packing order and strain}
\begin{figure}
\begin{center}
\includegraphics[width=0.4\textwidth]{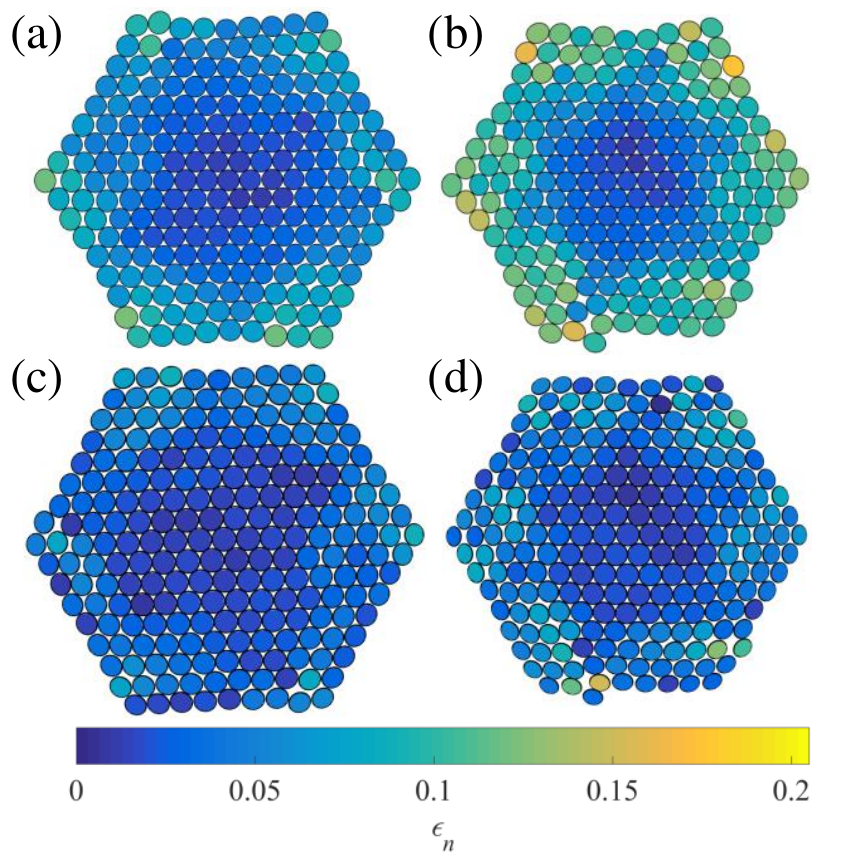}
\end{center}
\caption{Planar transect of the bundle corresponding to (a) $\theta_{max} = 20^\circ$  and (b) $\theta_{max} = 30^\circ$ in a plane perpendicular to the bundle axis labeled according to the strain $\epsilon_{n}$. (c,d) The corresponding disk packing on the dual surface given by Eq.~\ref{mapping} viewed from above. $\epsilon_{n}$ is observed to be significantly lower by comparison. 
}
\label{fig:coopmotion}
\end{figure}

An illustration of the discrepancy between the distance representation of the planar cut of the bundle and of the mapped packing on the curved surface, can be made through a comparison of the mean strain of inter-filament distances.  We define a nearest neighbor strain $\epsilon_{n} = (\langle\Delta \rangle - 1)$, where $\langle .. \rangle$ corresponding to averaging over the nearest neighbors, and $\Delta$ is computed either using $\Delta_{pp}$, or instead, on the curved surface described by Eq. (\ref{mapping}), using $\Delta_{dd}$. 
 We compare $\epsilon_{n}$ in the planar transect in Fig.~\ref{fig:coopmotion}(a,b) with those on the non-Euclidean surface in Fig.~\ref{fig:coopmotion}(c,d). A 3D view can be found in the corresponding animated GIFs included in the Supplementary Documentation.  
 
 While this mapping onto the surface shows that the filaments remain more-tightly packed than might appear while viewing a planar transect, it can be nonetheless noted that true inter-filament strains, develop increasingly in the outer layers of the filament bundles.  One observes in both cases that the strains predicted from filament positions in the 2D planar cut (Fig.~\ref{fig:coopmotion}a-b) consistently overestimate the inter-filament strains measured in 3D, or equivalently, between the mapped positions on the non-Euclidean surface (Fig.~\ref{fig:coopmotion}c-d).   These strains are the unavoidable consequence of {\it geometric frustration} encountered when packing lattices on surfaces with non-zero Gaussian curvature~\cite{rubinstein83,bowick09}, and equivalently in twisted bundles.

To understand the nature of frustration on positively-curved (e.g. spherical) surfaces, consider an annular ring of elements a distance $r$ from the center of an initially flat 2D lattice, which contains roughly $dN(r) \simeq 2\pi \rho_0~dr$ elements, where $\rho_0$ is the areal density.  Forcing the lattice onto a spherical surface of radius $R_s$ while maintaining the same distance from the center leads to a reduction of the perimeter of the annulus (the latitude at $r$) to $\ell(r) \simeq 2 \pi r(1-r^2/6R_s^2)$.  Hence, if the same $dN(r)$ are forces to occupy this annulus, the lattice is put under a compression in the hoop direction, and one which grows in magnitude with the distance from the center as $\sim (r/R_s)^2$.  If the lattice elements resist compression, like the nominally incompressible cross-section of filaments in these experiments, then some expansion outward is required to remove, or mitigate overlaps.  This is outward expansion, or aspiration, is precisely what is observed in the cross-section of bundles in Figs. \ref{fig:experimental}(c) and (d), where outer filament radii expand by 
6\% and 
9\%, for $\theta_{max}= 20^\circ$ and $30^\circ$, respectively.  Thus, the incompressible lattice packing responds to imposing of twist, by an outward expansion removes that overlaps along the hoop direction, at the expense of breaking contacts between radially-separated neighbors, a deformation pattern visible in the ``true strain" maps of  Fig.~\ref{fig:coopmotion}(c),(d).

While low to modest twist bundles result in relatively distributed patterns of inter-filament strain (as in Fig.~\ref{fig:coopmotion}c), larger twists ultimately disrupt the quasi-triangular packing, leading to the formation of localized defects at the bundle exterior.  Previous theory~\cite{amir12} and simulations~\cite{wang15} of the inter-filament stresses in twisted cohesive bundles, as well as the analogous problem of crystalline order on spherical caps~\cite{irvine10,amir14}, has shown that large-$N$ groundstates favor a similar development of lattice defects with dislocations decorating their outer compressive regions of the bundle above a critical twist.  In the present athermal system defect motion is expected to be kinetically inhibited (e.g. by the Peierls barrier~\cite{hirth}). Nonetheless, the localized deformations highlighted by large strains in Fig.~\ref{fig:coopmotion}(d) may then be evidence of incipient dislocations gliding in from the bundle surface to relax geometrically imposed compression, a possible mode of bundle plasticity.

\section{Conclusions}

We find that the structure of twisted elastic observed is consistent with a theoretical approach which postulates the equivalence between a packing of twisted filaments and a packing of disks on a non-Euclidean surface, a valuable tool for understanding the complex local and global constraints of packing imposed by twist. Our results further show the robustness of the approach because the filaments in the experiments have finite elasticity and residual friction.  Perhaps most non trivial is the fact that the approach applies to bundles that are held together and twisted mechanically only at the their ends, and yet naturally, adopt the constant-pitch configurations underlying the strict mapping from 3D spacing to a 2D non-Euclidean surface.  Applying the mapping to the experimental data, we were able to calculate the pattern of inter-filament strains and show that the core of the bundle largely preserves the hexagonal symmetry, which is less apparent from deformation pattern of the planar transect.  

Further, by examining the constraints imposed by the incompressibility of the filaments, we are able to explain the resulting expansion of the outer bundle with increasing twist.  This study gives the first experimental evidence of how the non-trivial constraints contact between extended and flexible elements lead to new responses associated with imposition on geometric frustration into an initially unfrustrated packing.   Future work will build on this work to develop and analyze the precise nature of the collective response of the initially unfrustrated packing to the imposition of geometric frustration through bundle twist, and further the mechanical work needed to bend filaments and reorganize their packing upon twist.
 
\begin{acknowledgments}
This work was supported by the National Science Foundation under grant number DMR 1508186 (AP, AK) and DMR 1608862 (GMG).  We thank D. Hall for valuable discussions.
\end{acknowledgments}





%
\bibliographystyle{unsrt}

\end{document}